\documentclass[showpacs,aps,floatfix,prb,reprint,superscriptaddress,showkeys]{revtex4-1}
\usepackage{xfrac}
\usepackage{amssymb}
\usepackage{braket}
\usepackage{graphicx}
\usepackage{verbatim}
\usepackage{etoolbox}
\usepackage{amsfonts}
\usepackage{amsmath}
\usepackage[utf8]{inputenc}
\usepackage{mathtools}
\usepackage{soul,xcolor,colortbl}
\usepackage{hyperref} \hypersetup{colorlinks=true,citecolor=blue,linkcolor=blue,urlcolor=black}
\usepackage{morefloats}
\usepackage{natmove}

\usepackage{bm}
\usepackage{array}
\newcolumntype{C}[1]{>{\centering\arraybackslash}p{#1}}\usepackage{soul}
\usepackage{color}
\def\AFLOW{{\small AFLOW}}
\def\AFLOWorg{{\sf \AFLOW.org}}

\def\AFLOWML{{\small AFLOW-ML}}
\def\GFA{{\small GFA}}
\def\JSON{{\small JSON}}
\def\API{{\small API}}
\def\ML{{\small ML}}

\def\citeAFLOWshort{\cite{aflowPAPER,curtarolo:art58,curtarolo:art104,aflowPI}}
\def\citeAFLOWLIB{\cite{aflowlibPAPER,curtarolo:art92,curtarolo:art104,aflux}}

\makeatletter \renewcommand\frontmatter@abstractwidth{\dimexpr\textwidth\relax} \makeatother

\begin{document}
\title{\Large Autonomous data-driven design of inorganic materials with AFLOW}

\author{Corey Oses}
\email[]{corey.oses@gmail.com}
\affiliation{Department of Mechanical Engineering and Materials Science, Duke University, Durham, North Carolina 27708, USA}
\author{Cormac Toher}
\email[]{toherc@gmail.com}
\affiliation{Department of Mechanical Engineering and Materials Science, Duke University, Durham, North Carolina 27708, USA}
\author{Stefano Curtarolo}
\email[]{stefano@duke.edu}
\affiliation{Materials Science, Electrical Engineering, Physics and Chemistry, Duke University, Durham NC, 27708}
\affiliation{Fritz-Haber-Institut der Max-Planck-Gesellschaft, 14195 Berlin-Dahlem, Germany}

\date{\today}

\begin{abstract}
\noindent
The expansion of programmatically-accessible materials data has cultivated opportunities
for data-driven approaches.
Highly-automated frameworks like \AFLOW\ not only manage the generation, storage, and dissemination of materials data,
but also leverage the information for thermodynamic formability modeling,
such as the prediction of phase diagrams and properties of disordered materials.
In combination with standardized parameter sets, the wealth of data
is ideal for training machine learning algorithms, which
have already been employed for property prediction,
descriptor development, design rule discovery, and the identification of candidate functional materials.
These methods promise to revolutionize the path to synthesis and, ultimately, transform
the practice of traditional materials discovery to one of rational and autonomous materials design.
\end{abstract}

\maketitle

\section*{Introduction}
Density functional theory implementations~\cite{kresse_vasp,VASP4_2,vasp_cms1996,vasp_prb1996,quantum_espresso_2009,gonze:abinit,Blum_CPC2009_AIM}
offer a reasonable compromise between cost and accuracy~\cite{Haas_PRB_2009},
stimulating rapid development of automated frameworks and corresponding data repositories.
Prominent examples include
\AFLOW\ (the \underline{A}utomatic \underline{Flow} Framework for Materials Discovery)~\citeAFLOWshort,
{\small NoMaD} (\underline{No}vel \underline{Ma}terials \underline{D}iscovery Laboratory)~\cite{nomad},
Materials Project~\cite{APL_Mater_Jain2013},
{\small OQMD} (\underline{O}pen \underline{Q}uantum \underline{M}aterials \underline{D}atabase)~\cite{Saal_JOM_2013},
Computational Materials Repository and its associated scripting interface {\small ASE} (\underline{A}tomic \underline{S}imulation \underline{E}nvironment)~\cite{cmr_repository},
and {\small AiiDA} (\underline{A}utomated \underline{I}nteractive \underline{I}nfrastructure and \underline{Da}tabase for Computational Science)~\cite{Pizzi_AiiDA_2016}.
Such repositories house an abundance of materials data.
For instance, the \AFLOWorg\ database contains
over 1.8 million compounds, each characterized by about 100 different properties~\cite{aflowlibPAPER,curtarolo:art92,curtarolo:art104,aflux}.
Investigations employing this data have not only led to advancements in modeling
electronics~\cite{curtarolo:art77,curtarolo:art94,curtarolo:art124,ceder:nature_1998},
thermoelectrics~\cite{curtarolo:art120,curtarolo:art129},
superalloys~\cite{curtarolo:art113},
and metallic glasses~\cite{curtarolo:art112},
but also the synthesis of two new magnets --- the first
discovered by computational approaches~\cite{curtarolo:art109}.

Further advancements and discoveries are contingent on continued development and expansion of these materials repositories.
New entries are generated both by \textbf{i.} calculating the properties of previously observed compounds
from sources such as the Inorganic Crystal Structure Database~\cite{ICSD},
and \textbf{ii.} decorating structural prototypes~\cite{aflowANRL} to predict new materials.
Accurate computation of materials properties --- including electronic,
magnetic, chemical, crystallographic, thermomechanical, and thermodynamic features ---
demands a combination of
\textbf{i.} reliable calculation parameters/thresholds~\cite{curtarolo:art104} and
\textbf{ii.} robust algorithms that scale with the size/diversity of the database.
For example, convenient definitions for the primitive cell representation~\cite{aflowPAPER} and
high-symmetry Brillouin Zone path~\cite{curtarolo:art58} have both optimized and standardized electronic structure calculations.
Moreover, careful treatment of spatial tolerance and proper validation schemes have finally
enabled accurate and fully autonomous determination of the
complete symmetry profile of crystals~\cite{curtarolo:art135},
which is essential for elasticity~\cite{curtarolo:art115} and phonon~\cite{aflowPAPER,curtarolo:art114,curtarolo:art119,curtarolo:art125}
calculations.

Beyond descriptions of simple crystals,
exploration of complex properties~\cite{curtarolo:art96,curtarolo:art115} and materials~\cite{curtarolo:art110,curtarolo:art112}
typically warrants advanced (and expensive) characterization techniques~\cite{Hedin_GW_1965,GW,ScUJ}.
Fortunately, state-of-the-art workflows~\cite{curtarolo:art96,curtarolo:art110,curtarolo:art115} and
careful descriptor development~\cite{curtarolo:art112} have
enabled experimentally-validated modeling within a density functional theory framework.
Furthermore, the combination of plentiful and diverse materials data~\citeAFLOWLIB\
and its programmatic accessibility~\cite{curtarolo:art92,aflux}
justify the application of data-mining techniques.
These methods can quantitatively resolve subtle trends and correlations among
materials and their properties~\cite{curtarolo:art94,Ghiringhelli_PRL_2015,curtarolo:art124,curtarolo:art129,curtarolo:art135},
as well as motivate the formulation of novel property descriptors~\cite{curtarolo:art112,Lederer_HEA_2018}.
These ``black-box'' models are surprisingly accurate and quite valuable, particularly
when few practical alternative modeling schemes exist
--- as is the case for predicting superconducting critical temperatures~\cite{curtarolo:art94,curtarolo:art137}.

Ultimately, the power in \ML\ lies in the speed of its predictions, which out-paces
density functional theory calculations by orders of magnitude~\cite{Isayev_ChemSci_2017}.
Given that the number of currently characterized materials pales in comparison to the full space of hypothetical structures,
methods to filter/screen the most interesting candidate materials~\cite{Walsh_NChem_2015} --- powered by \ML\ models --- will
undoubtedly become integral to future materials discovery workflows.

\section*{Thermodynamic formability modeling}

\begin{figure*}
  \includegraphics[width=1.00\textwidth]{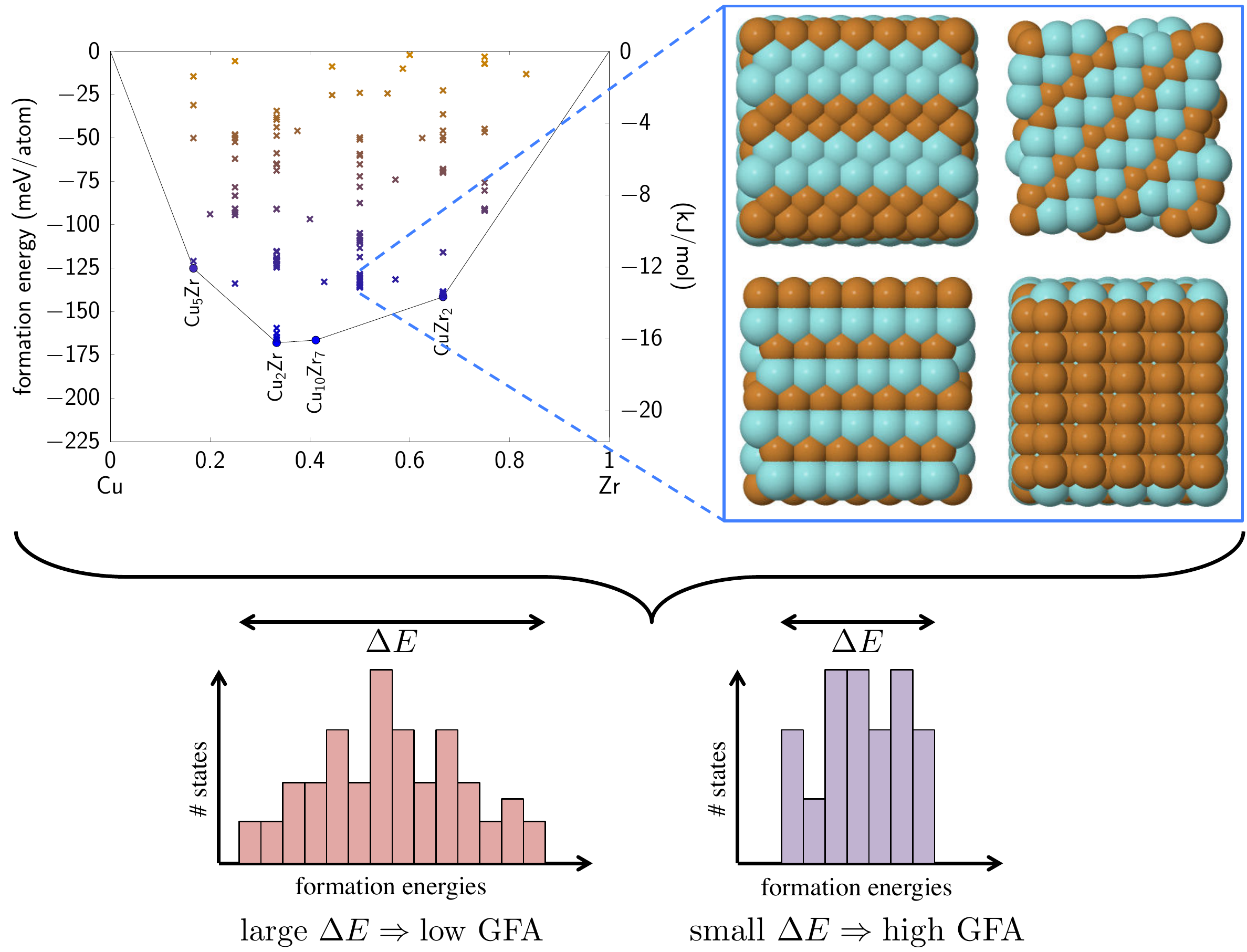}
  \vspace{-4mm}
  \caption{\small \textbf{Descriptor for glass forming ability.}
    The glass forming ability of metal alloy systems can be predicted from the
    spread of formation energies of relevant ordered structures.
    Different structural phases with similar energies compete against each other during solidification,
    frustrating crystallization and promoting glass formation.
    A broad distribution of energies implies a low glass forming ability,
    while a narrow distribution indicates a high glass forming ability.}
  \label{fig:GFA_descriptor}
\end{figure*}

\noindent\textbf{Prediction of phase diagrams.}
Descriptions of thermodynamic stability and structural/chemical disorder are resolved
through statistical analyses of aggregate sets of structures.
Thermodynamic stability largely governs synthesizability, which can be determined by an
analysis of how structures of similar compositions compete energetically, \textit{i.e.},
determination of the minimum Gibbs free energy surface.
The procedure is algorithmically equivalent to finding the lower-half convex hull of all the relative
free energy minima~\cite{curtarolo:art20} as illustrated in Figure~\ref{fig:GFA_descriptor}.
Composition and energy information from relevant \AFLOWorg\ calculations are plotted,
and the phases defining the minimum energy surface are identified~\cite{qhull}.
Assuming sufficient sampling, the ground state structures on the minimum energy surface
form the low-temperature phase diagram~\cite{monsterPGM}.

The convex hull construction offers a wealth of related thermodynamic properties.
For near-hull structures, the energetic distance from the minimum energy surface
is treated as a metric for synthesizability, as only small perturbations
in temperature or pressure may be needed for it to be realized.
In fact, this distance is equivalent to the amount of energy
driving the decomposition of an unstable state to a linear combination
of nearby ground state structures.
A similar distance --- that of a stable phase from the pseudo-hull formed by
neglecting it --- quantifies
the impact of a structure on the minimum energy surface and characterizes
the robustness of stable structures, \textit{i.e.}, the stability criterion.

\AFLOW\ offers a module for autonomous calculation of the convex hull,
which retrieves the set of relevant structure
calculations from the repository~\cite{curtarolo:art92,aflux}
and delivers a thorough thermodynamic characterization for each.
Filtering schemes based on these thermodynamic properties, including the stability criterion and
tie-line construction, played key roles in the discovery of new magnets~\cite{curtarolo:art109}
and modeling superalloys~\cite{curtarolo:art113}.
The module powers an online web application for enhanced visualization of two/three-dimensional hulls
available at {\sf aflow.org/aflow-chull}.

\noindent\textbf{Modeling disordered materials.}
Incorporating the effects of disorder is a necessary, albeit difficult, step in materials modeling.
Not only is disorder intrinsic to all materials,
but it also offers a route to enhanced and even otherwise inaccessible functionality.
Disordered materials range from chemically disordered high entropy materials and solid solutions,
in which sites on a periodic crystal lattice are randomly occupied,
to structurally disordered amorphous glasses, with no crystalline periodicity.
Materials such as high entropy alloys~\cite{Gao_HEA_book_2015, Miracle_HEAs_NComm_2015}
containing four to five metallic elements in equi-composition are being investigated
for their enhanced thermomechanical
properties~\cite{HEAapp1,MoNbTaW,Gludovatz_hea_mech_properties,MoNbTaW2,HEAprop2},
and have also been reported to display superconductivity~\cite{HEAprop1}.
Research interest has recently expanded beyond metallic alloys to include high entropy
ceramics such as entropy stabilized oxides~\cite{curtarolo:art99,curtarolo:art122} and
high-entropy borides~\cite{Gild_borides_SciRep_2016}, which display
promising behavior including colossal dielectric constants~\cite{Beradan_2016_PSSA_ESO_Colossal}
and superionic conductivity~\cite{Beradan_2016_JMCA_ESO_superionic}.

\textit{Ab-initio} modeling of chemical/substitutional disorder --- including
vacancies and random site occupations --- is a notoriously formidable problem,
since it results in systems that cannot be described directly by a single unit cell with periodic boundary conditions.
Rigorous statistical treatment of chemical disorder leverages
a set of representative ordered supercells in thermodynamic competition.
System-wide properties are resolved through ensemble averages of these supercells.
The approach has been implemented in \AFLOW~\cite{curtarolo:art110} for autonomous
characterization, and successfully validated for a number of technologically significant systems,
recovering characteristic trends as a function of composition and offering
additional insight into underlying physical mechanisms.
The module determines the smallest superlattice size that
accommodates the required stoichiometry to within a user-defined tolerance,
and then generates the corresponding superlattices using Hermite Normal Form matrices~\cite{gus_enum1}.
All allowed decoration permutations are considered for each superlattice variant, generating the
full set of possible supercell configurations.
Degeneracies are rapidly identified by comparing
approximate structure energies calculated with the Universal Force Field method~\cite{Rappe_1992_JCAS_UFF}.
Only unique supercells are individually characterized using standard
\textit{ab-initio} packages~\cite{kresse_vasp,VASP4_2,vasp_cms1996,vasp_prb1996,quantum_espresso_2009,gonze:abinit,Blum_CPC2009_AIM}.
The ensemble average values of properties such as the electronic band gap, density of states, and the magnetic moment
--- weighted according to a Boltzmann distribution for a particular temperature --- are
then calculated to resolve the behavior of the disordered material.

\begin{figure*}
  \includegraphics[width=1.00\textwidth]{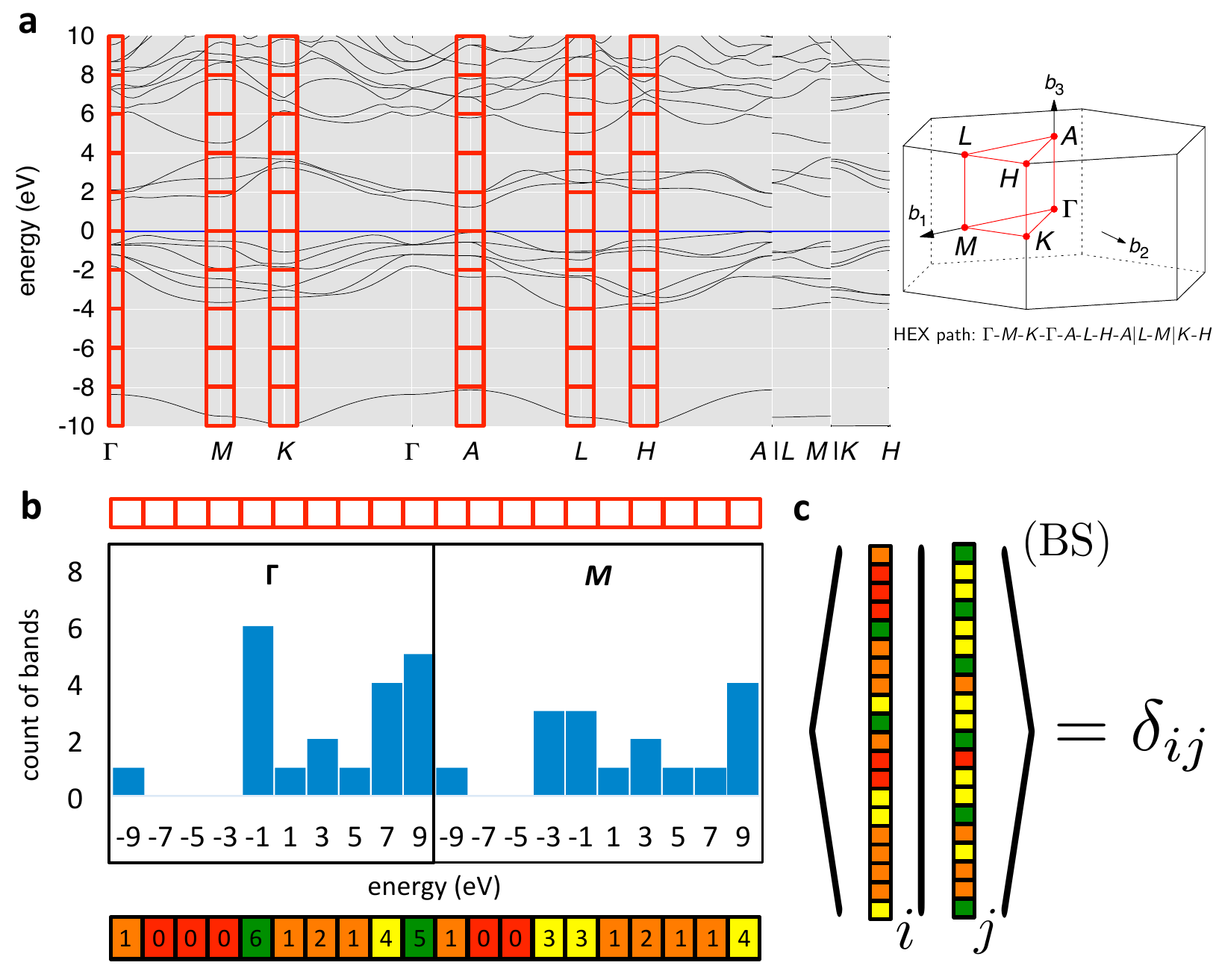}
  \vspace{-4mm}
  \caption{\small \textbf{Construction of electronic structure fingerprints.}
    (\textbf{a}) The density of electronic states in certain energy ranges at the high-symmetry points of the Brillouin Zone
    (\textbf{b}) form a fingerprint for the structure.
    (\textbf{c}) The fingerprints for different materials are then compared to quantify their similarity.}
  \label{fig:ML_fingerprints}
\end{figure*}

Metallic glasses lack an ordered lattice, and its associated defects, which endow them
with a unique combination of superb mechanical properties~\cite{chen2015does} and
plastic-like processability~\cite{schroers2006amorphous,Schroers_blow_molding_2011,kaltenboeck2016shaping},
rendering them of great interest for several potential commercial and industrial
applications~\cite{Schroers_Processing_BMG_2010,johnson2016quantifying,ashby2006metallic}.
To predict the \underline{g}lass \underline{f}orming \underline{a}bility (\GFA)
of metal alloy systems~\cite{curtarolo:art112},
statistical approaches have been employed that blend the concept of thermodynamically competing ordered structures with
the large quantities of pre-calculated data available in the \AFLOWorg\ repository.
The proposed physical mechanism is that ordered phases which have similar energies,
but are structurally distinct, will compete against each other during solidification,
frustrating crystal nucleation and thus promoting glass formation, as illustrated in Figure \ref{fig:GFA_descriptor}.
The energy distribution of the different structures can be considered as forming a thermodynamic density of states:
a narrow distribution indicates a high \GFA, while a wider distribution implies a low \GFA.
Atomic environment~\cite{daams_villars:environments_2000,daams:cubic_environments}
comparisons determine the similarity of ordered crystalline phases,
enabling the formulation of a quantitative descriptor that can be applied to the entire \AFLOWorg\ database.
The different structures are weighted according to a Boltzmann distribution to create the \GFA\ descriptor.
The model is found to successfully predict 73\% of the glass forming compositions
for a set of 16 experimentally well-characterized alloy systems, and also indicates that about 17\% of binary alloy systems
should be capable of glassification.
By exploiting the pre-calculated data in the \AFLOWorg\ repository, this model can be leveraged to rapidly
predict \GFA\ as a function of composition for thousands of alloy systems,
demonstrating the power of applying intelligently constructed descriptors to computational materials data.

The \AFLOW\ formation energy data is also employed to train cluster expansion models~\cite{atat2},
which can be combined with thermodynamic modeling to predict the order-disorder transition temperature for
solid-solutions in high-entropy alloys~\cite{Lederer_HEA_2018}.
Order-disorder transitions in the form of spinodal decomposition have also been proposed as a mechanism
to \textbf{i.} embed topologically-protected conducting interface states in an insulating matrix~\cite{curtarolo:art134}
and \textbf{ii.} self-assemble nanostructures (such as thermoelectric devices~\cite{curtarolo:art107}).
The boundaries between different layers act as phonon-scatterers, suppressing the thermal conductivity and thus improving efficiency.

\section*{Exploiting machine learning algorithms}

\noindent\textbf{Model development.}
\ML\ is rapidly emerging as a powerful tool for computational materials
design~\cite{Bhadeshia_ISIJ_ML_1999,Ghiringhelli_PRL_2015,PyzerKnapp_AdFM_2015,Guzik_NMat_2016}.
Given sufficient training data, algorithms such as
neural networks~\cite{sumpter_nnetworks_review_1996},
random forests~\cite{randomforests},
gradient boosting decision trees~\cite{Friedman_AnnStat_2001}
and support vector machines~\cite{Cortes_ML_SVM_1995}
can learn to
\textbf{i.} identify the structures that are thermodynamically accessible for a given composition~\cite{Ghiringhelli_PRL_2015}
and
\textbf{ii.} accurately predict materials properties, such as the
electronic band gap~\cite{curtarolo:art124},
elastic moduli~\cite{curtarolo:art124, deJong_SR_2016},
vibrational energies~\cite{curtarolo:art129}, and
lattice thermal conductivity~\cite{curtarolo:art84}.

The successful training of \ML\ models depends crucially on the set of features
characterizing the material, \textit{i.e.}, the set of descriptors that form the feature vector~\cite{Ghiringhelli_PRL_2015}.
Such representations include electronic structure fingerprints~\cite{curtarolo:art94}
and crystal graphs~\cite{curtarolo:art124,Xie_ML_CNN_2017}.
Optimal descriptors are resolved by exploring different
linear and non-linear combinations of properties, and extracting the most
efficient feature vector via compressive sensing~\cite{curtarolo:art135}.
Compressive sensing finds the sparse solution ($\ell_0$-norm minimization) of the
underdetermined system of linear equations
mapping the set of observable materials properties to the large set of
possible test features --- effectively reducing the dimensionality of the problem.
The algorithm also filters for physically meaningful combinations of properties, based
on dimensional analysis, to maximize interpretability of the final descriptor set.

Several different \ML\ frameworks are leveraging data from the \AFLOWorg\ repository.
The materials fingerprinting model~\cite{curtarolo:art94} codifies
aspects of the electronic structure~\cite{curtarolo:art58}
to serve as unique markers for each material.
In particular, the number of bands that intersect high-symmetry Brillouin Zone points
at discretized energy values form the band structure fingerprint (illustrated in Figure~\ref{fig:ML_fingerprints}),
while simple discretization of the density of states form the density of states fingerprint.
The Tanimoto coefficient --- a distance metric~\cite{Bajusz_JCheminfo_2015} --- between
fingerprint vectors quantifies the similarity of the electronic structure between different materials.
These fingerprints are employed for the construction of networks, \textit{i.e.},
materials cartography~\cite{curtarolo:art94}, where materials are represented by nodes
and similarity correlates with relative positioning.
When applied to compounds in the Inorganic Crystal Structure Database,
significant clustering and structure can be identified for these networks,
particularly with respect to material complexity (binaries \textit{versus} ternaries, \textit{etc.}), type (metal \textit{versus} insulator),
and, surprisingly, superconducting critical temperature~\cite{curtarolo:art94}.

In the case of high-temperature superconductors, significant clustering suggests
strong correlations among the electronic structure of these materials; although, as expected,
these features alone are not enough to quantitatively resolve critical temperatures.
Indeed, modeling improves with integration of more experimental observations~\cite{Supercon,curtarolo:art137}
and properties, such as structural features and partial charges~\cite{bader3}.
Incorporating additional relevant and physically meaningful training data, such
as the phonon spectra, should offer an applicability domain expansion
and higher fidelity predictions.

Thermomechanical properties calculated using the elastic constants~\cite{curtarolo:art115} and
Debye-Gr{\"u}neisen~\cite{curtarolo:art96} modules of \AFLOW\ have been employed to train
a gradient boosting decision trees framework~\cite{curtarolo:art124} to predict quantities
such as the bulk and shear modulus, Debye temperature, and heat capacity.
Indicative of its versatility, the same model~\cite{curtarolo:art124} has also been trained on \AFLOW\ electronic structure
data to classify materials as metals or insulators, and to predict the electronic band gap for compounds
identified as non-metals.
Model development is based on a fragment construction approach: each crystal
is represented by a graph where nodes are decorated with corresponding atomic properties
and connectivity is dictated by distance and Voronoi polyhedra adjacency.
Path and circular fragments --- representative of linear geometry and coordination polyhedra
within the crystal --- form the basis for feature development.
To train the models, the gradient boosting decision tree algorithm is employed,
which amalgamates a series of weak, easily constructed prediction rules
to resolve a single, highly predictive function.

The resulting models have been thoroughly validated with simulated and real tests sets,
showing predictive metrics at 90\% or higher against existing calculated
and experimental measurements.
Beyond property value prediction, feature-importance analyses of the models recovers
meaningful ways to tune the band gap and Debye temperature, offering
practical design rules for device engineering.
The development of such models achieves the greatest impact on thermomechanical properties,
where characterizations demand many single \textit{ab-initio} calculations,
and thus presents a substantial boost in prediction speed at a fraction of the resources.

\noindent\textbf{Workflow integration.}
\ML\ approaches are expected to become indispensable in two specific scenarios, prediction
of complex properties and screening of large sets of materials.
Unfortunately, widespread exploitation of \ML\ techniques in materials science has been
hindered by the difficulty of setting up and interfacing the models with materials design infrastructures.
To streamline this process, the \AFLOWML\ \API~\cite{aflowmlapi}
has been created to provide programmatic access to the \ML\ models described in
Refs.~\onlinecite{curtarolo:art124} and \onlinecite{curtarolo:art129},
with plans to extend it with additional models as they are developed.
By posting a structure file to the \API, users can retrieve \ML\ predictions of thermal,
mechanical, and electronic properties in the \JSON\ data format.
In this way, all technical details of the \ML\ algorithms are abstracted away, rendering
a simple interface no more complicated than that of a standard \API.
This procedure can be easily incorporated into materials design workflows,
due to its use of ubiquitous HTTP commands, along with the \JSON\ format that is easily
interpreted by a wide range of modern programming languages.

\section*{Source code and online forum}
Since 2018, the \AFLOW\ software (V3.1.193) has been made available for
download/redistribution under the terms of the GNU License {\sf http://www.gnu.org/licenses}.
The source code/license/readme files can be found by following the links in the \AFLOWorg\ project website.
Though some of the aforementioned modules are conveniently interfaced
through the website, only the executable offers full and unabridged functionality.
Additionally, the Forum ({\sf aflow.org/forum}) advertises updates and new functionality,
as well as hosts discussion boards for registered members to post questions. \\

\section*{Conclusion}
Broad scale thermodynamic formability modeling and exploitation of \ML\ algorithms
represent current frontiers in computational materials design.
Recent progress in these fields has been enabled by large, programmatically-accessible materials
databases generated by automated computational infrastructures.
Ensembles of ordered phases are being successfully employed to \textbf{i.} construct phase diagrams and
\textbf{ii.} formulate descriptors and models to predict the formation and properties of disordered materials.
\ML\ models have the potential to rapidly accelerate materials design as tools for
predicting properties and identifying subtle/hidden trends --- thus leading to enhanced
understanding of the physical mechanisms underlying materials behavior.

\section*{Acknowledgments}
We thank Drs. E. Perim, Y. Lederer, O. Levy, O. Isayev, A. Tropsha, N. Mingo,  J. Carrete,
J.~J. Vlassak, J. Schroers, D. Hicks, and E. Gossett for insightful discussions.
CO acknowledges support from the NSF Graduate Research Fellowship \#DGF1106401.
SC acknowledges support by the Alexander von Humboldt-Foundation.

\section*{Corresponding authors}
Correspondence should be directed to
Corey Oses (\verb|corey.oses@gmail.com|),
Cormac Toher (\verb|toherc@gmail.com|),
and/or Stefano Curtarolo (\verb|stefano@duke.edu|).

\newcommand{\Ozolins}{Ozoli\c{n}\v{s}}

\end{document}